\documentclass[notitlepage,aps,pra,twoside,superscriptaddress,showkeys,nofootinbib]{revtex4-1}
\usepackage{graphicx}
\usepackage{amsmath,amsfonts,amssymb,amsthm,amsbsy,mathtools}
\usepackage{bbold}
\usepackage{epic}
\usepackage{eepic}
\usepackage{mathrsfs}
\usepackage[all]{xy}
\usepackage{lmodern}
\usepackage[T1]{fontenc}
\usepackage[utf8]{inputenc}
\usepackage{color}
\usepackage{xr}

\newcommand{\ket}[1]{\mathop{\left|#1\right>}\nolimits}            
\newcommand{\bra}[1]{\mathop{\left<#1\,\right|}\nolimits}         

\newcommand{\brk}[2]{\langle #1 | #2 \rangle}
\newcommand{\ketbra}[2]{| #1\rangle\!\langle #2 |}

\def\e{\epsilon}

\def\e{\mathrm{e}}

\newmuskip\pFqmuskip   

\newcommand*\pFq[6][8]{%
  \begingroup 
  \pFqmuskip=#1mu\relax
  \mathcode`\,=\string"8000
  \begingroup\lccode`\~=`\,
  \lowercase{\endgroup\let~}\pFqcomma
  {}_{#2}F_{#3}{\left[\genfrac..{0pt}{}{#4}{#5};#6\right]}
  \endgroup
}
\newcommand{\pFqcomma}{\mskip\pFqmuskip}

\begin{document}

\title{Entanglement of quantum clocks through gravity}

\author{Esteban Castro-Ruiz}
\author{Flaminia Giacomini}
\author{\v{C}aslav Brukner}
\affiliation{Vienna Center for Quantum Science and Technology (VCQ), Faculty of Physics,
University of Vienna, Boltzmanngasse 5, A-1090 Vienna, Austria}
\affiliation{Institute for Quantum Optics and Quantum Information (IQOQI),
Austrian Academy of Sciences, Boltzmanngasse 3, A-1090 Vienna, Austria}

\begin{abstract}
In general relativity, the picture of spacetime assigns an ideal clock to each world line. Being ideal, gravitational effects due to these clocks are ignored and the flow of time according to one clock is not affected by the presence of clocks along nearby world lines. However, if time is defined operationally, as a pointer position of a physical clock that obeys the principles of general relativity and quantum mechanics, such a picture is, at most, a convenient fiction. Specifically, we show that the general relativistic mass-energy equivalence implies gravitational interaction between the clocks, whereas the quantum mechanical superposition of energy eigenstates leads to a nonfixed metric background. Based only on the assumption that both principles hold in this situation, we show that the clocks necessarily get entangled through time dilation effect, which eventually leads to a loss of coherence of a single clock. Hence, the time as measured by a single clock is not well defined. However, the general relativistic notion of time is recovered in the classical limit of clocks.
\end{abstract}

\maketitle



\section{Introduction}
A crucial aspect of any physical theory is to describe the behaviour of systems with respect to the passage of time. Operationally, this means to establish a correlation between the system itself and another physical entity, which acts as a clock. In the context of general relativity, time is specified locally in terms of the proper time along world lines. It is believed that clocks along these world lines correlate to the metric field in such a way that their readings coincide with the proper time predicted by the theory -- the so-called ``clock hypothesis'' \cite{brown}. A common picture of a reference frame uses a latticework of clocks to locate events in spacetime \cite{taylor}. An observer, with a particular split of spacetime into space and time, places clocks locally, over a region of space. These clocks record the events and label them with the spatial coordinate of the clock nearest to the event and the time read by this clock when the event occurred. The observer then reads out the data recorded by the clocks at his/her location. Importantly, the observer does not need to be sitting next to the clock in order to do so. We will call an observer that measures time according to a given clock, but not located next to it, a far-away observer.

In the clock latticework picture, it is conventionally considered that the clocks are external objects which do not interact with the rest of the Universe. This assumption does not treat clocks and the rest of physical systems on equal footing and therefore is artificial. In the words of Einstein: ``One is struck [by the fact] that the theory [of special relativity]... introduces two kinds of physical things, i.e. (1) measuring rods and clocks, (2) all other things, e.g., the electromagnetic field, the material point, etc. This, in a certain sense, is inconsistent...''\cite{EinsteinAutobiographicalNotes}. For the sake of consistency, it is natural to assume that the clocks, being physical, behave according to the principles of our most fundamental physical theories: quantum mechanics and general relativity.

In general, the study of clocks as quantum systems in a relativistic context provides an important framework to investigate the limits of the measurability of spacetime intervals \cite{salecker}. Limitations to the measurability of time are also relevant in models of quantum gravity \cite{amelino-camelia, gambini}. It is an open question how quantum mechanical effects modify our conception of space and time and how the usual conception is obtained in the limit where quantum mechanical effects can be neglected.

In this work we show that quantum mechanical and gravitational properties of the clocks put fundamental limits to the joint measurability of time as given by clocks along nearby world lines. As a general feature, a quantum clock is a system in a superposition of energy eigenstates. Its precision, understood as the minimal time in which the state evolves into an orthogonal one, is inversely proportional to the energy difference between the eigenstates \cite{mandelstam,fleming,margolus,aharonov,busch}. Due to the mass-energy equivalence, gravitational effects arise from the energies corresponding to the state of the clock. These effects become non-negligible in the limit of high precision of time measurement. In fact, each energy eigenstate of the clock corresponds to a different gravitational field. Since the clock runs in a superposition of energy eigenstates, the gravitational field in its vicinity, and therefore the spacetime metric, are in a superposition. We show that, as a consequence of this fact, the time dilation of clocks evolving along nearby world lines is ill-defined. As we will see below, this effect is already present in the weak-gravity and slow velocities limit, in which the number of particles is conserved. Moreover, it leads to entanglement between nearby clocks, implying that there are fundamental limitations to the measurability of time as recorded by the clocks. 

The limitation, stemming from quantum mechanical and general relativistic considerations, is of a different nature than the ones in which the spacetime metric is assumed to be fixed \cite{salecker}. Other works regarding the lack of measurability of time due to the effects the clock itself has on spacetime \cite{amelino-camelia, gambini}, argue that the limitation arises from the creation of black holes. We will show that our effect is independent of this effect, too. Moreover, it is significant in a regime orders of magnitude before a black hole is created. Finally, we recover the classical notion of time measurement in the limit where the clocks are increasingly large quantum systems and the measurement precision is coarse enough not to reveal the quantum features of the system. In this way we show how the (classical) general relativistic notion of time dilation emerges from our model in terms of the average mass-energy of a gravitating quantum system.

From a methodological point of view, we propose a \textit{gedanken} experiment where both general relativistic time dilation effects and quantum superpositions of spacetimes play a significant role. Our intention, as is the case for \textit{gedanken} experiments, is to take distinctive features from known physical theories (quantum mechanics and general relativity, in this case) and explore their mutual consistency in a particular physical scenario. We believe, based on the role \textit{gedanken} experiments played in the early days of quantum mechanics and relativity, that such considerations can shed light on regimes for which there is no complete physical theory and can provide useful insights into the physical effects to be expected at regimes that are not within the reach of current experimental capabilities.    

\section{The clock model}
\label{clockmodel}
Any system which is in a superposition of energy eigenstates can be used as a reference clock with respect to which one defines time evolution. 
The simplest possible case is that in which the clock is a particle with an internal degree of freedom that forms a two-level system. 
In the following, we assume the clock to follow a semiclassical trajectory which is approximately static, that is, it has (approximately) zero velocity with respect to the observer that uses the clock to define operationally his/her reference frame, in the sense stated above. In this way, special-relativistic effects can be ignored. We stress the fact that the observer does not need to be located next to the clock. He/she can perform measurements on it by sending a probe quantum system to interact with the clock and then measuring the probe in his/her location. In the following, we focus only on the clock's internal degrees of freedom, which are the only ones relevant to our model. The internal Hamiltonian of the particle  in its rest reference frame,  
\begin{align}
\label{internal hamiltonian}
H_{int} = E_0 \ketbra{0}{0} + E_1 \ketbra{1}{1}, 
\end{align}
generates the evolution of the clock. For convenience we choose the origin of the energy scale so that $E_0 = 0$, and we define $\Delta E = E_1-E_0 = E_1$.

An operational meaning of the `passage of a unit of time', in which by definition the system goes through a noticeable change from an initial state to a final state, can be given in terms of the \textit{orthogonalisation time} of the clock, that is, the time it takes for the initial state to become orthogonal to itself. For a two level system, the orthogonalisation time is equal to
$t_\perp = \hbar \pi/\Delta E$ \cite{magdalena}. Note that $t_\perp$ quantifies the precision of the clock and it is in this sense a measure of time uncertainty. The optimal initial state of the clock is one with an equal superposition of energies, which we choose to be
\begin{equation}
\label{initial state 2}
\ket{\psi_{in}} = \frac{1}{\sqrt{2}}(\ket{0}+\ket{1}).
\end{equation}
For this state, the optimal measurement to determine the passage of time is given by projectors in the $\ket{\pm} = (\ket{0}\pm\ket{1})/\sqrt{2}$ basis. 
It is important to stress that the relation between orthogonalisation time and energy difference is fundamental: any clock model has a precision limited by the difference of energies involved in the time measuring process. This fact was already noticed in earlier works \cite{salecker,aharonov}. It is this feature, also shared by more detailed clock models \cite{buzek}, which plays a fundamental role in this work. The fact that the clock can return periodically to its initial state and therefore give ambiguous time readings can be dealt with by choosing a more elaborate clock model, e.g. a system with more energy levels. This fact is irrelevant for the result of this section and hence we will treat here only the two level case. This two-level clock model does not aim to describe all the features involved in time measurements, like for example the reconstruction of the `flow of time' from repetitions of measurements \cite{salecker,rankovic}. Our intention in this section is to point out the \textit{minimal requirements} for a system to be a clock, i.e. that the system must be in a superposition of energy eigenstates. It follows from these requirements that the orthogonalisation time is inversely proportional to the energy gap of the clock. A more elaborate model of a clock, that addresses the issue of repetitive measurements, will be considered below, when studying how the general relativistic notion of time dilation emerges in the classical limit. 

\begin{figure}[h]
\centering
\includegraphics[scale=0.25]{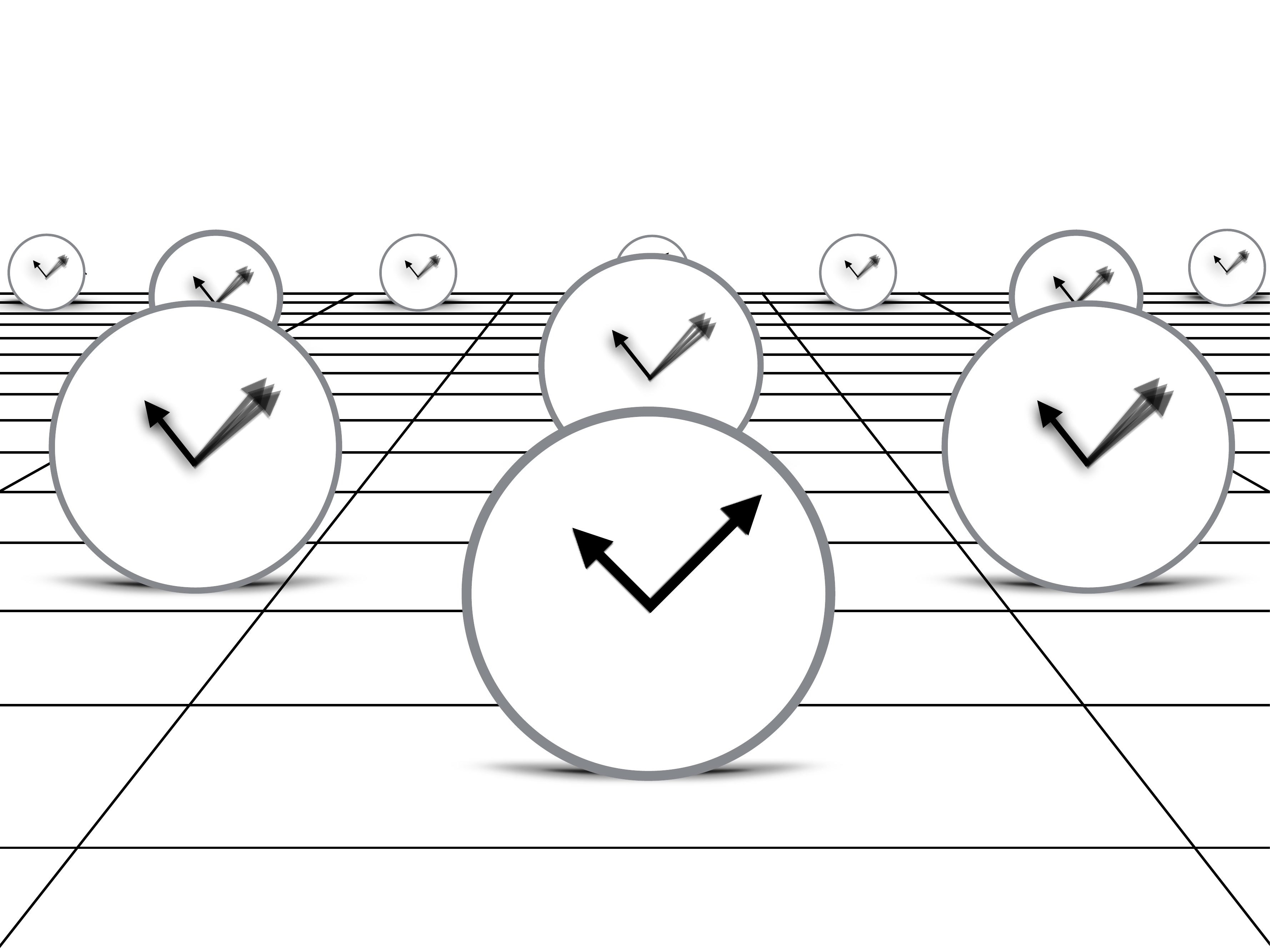}
\caption{\label{clocksonperspective}Pictorial representation of the fundamental trade-off between uncertainty of time measurement by a given clock and uncertainty of time measurement by nearby clocks. The clock at the frontal plane of the picture has a relatively high accuracy, depicted by its sharply defined hands. The uncertainty of time-reading for this clock is inversely proportional to the energy gap $\Delta E$ of the internal degree of freedom that constitutes the clock (see Section \ref{clockmodel} in the main text). By the mass-energy equivalence, the energy of the clock will produce gravitational time-dilation effects on nearby clocks. Since the energy is not well-defined but has an uncertainty $\Delta E$, nearby clocks will have an uncertainty in their time dilation with respect to the main clock, as depicted by the `fuzzy' hands in a superposition. There exists therefore a limitation to the possibility of defining time accurately at nearby points, given by the joint effects of quantum mechanics (superposition principle) and general relativity (gravitational time dilation). This effect is fundamental and independent of the energy gap $\Delta E$ of the clock, as stated in Eq. (\ref{clocks uncertainty relation}).}
\end{figure}

The gravitational effects due to the energies involved are to be expected at a fundamental level. In particular, for a given energy of the clock, there is a time dilation effect in its surroundings, due to the mass-energy equivalence. However, since the mass-energy corresponding to the amplitude of $\ket{0}$ is different to that corresponding to $\ket{1}$, the time dilation in the vicinity of the clock in state given by Eq. (\ref{initial state 2}) is uncertain, see Fig. (\ref{clocksonperspective}). Consider a second clock localised at a coordinate distance $x$ from the first clock (in the reference frame of the before mentioned observer). Due to time dilation, this clock would run as in flat spacetime for the amplitude corresponding to $\ket{0}$, and it would run (to second order approximation in $c^{-2}$) as  
$t\longrightarrow t+\Delta t = t\left(1+G\Delta E/(c^4 x)\right)$,
for the amplitude corresponding to the excited state $\ket{1}$. Here, $G$ denotes the gravitational constant and $t$ can be operationally defined as the proper time of the observer, who is sufficiently far away from the mass-energy distribution so that the effects of the different gravitational fields originating from the two states of the clocks are indistinguishable at his/her location. This observer ensures that the coordinate distance $x$ between the first observer and the clocks is kept fixed. In Appendix (A) we quantify the minimum distance between the observer and the clocks such that he/she cannot operationally distinguish between the different gravitational fields.

As a consequence of these considerations, there is a fundamental trade-off between the accuracy of measuring time at the location of the clock and the uncertainty of time dilation at nearby points. It can be succinctly described by the relation
\begin{equation}
\label{clocks uncertainty relation}
t_\perp \Delta t = \frac{\pi \hbar G t}{c^4 x},
\end{equation}
which is an uncertainty relation that arises due to both quantum mechanical and general relativistic effects. It holds independently of the energy gap of the clock or its particular constitution. 

So far, our treatment of time dilation in the vicinity of the clock has been classical and non-operational. In the following, we explain the above effect in terms of gravitational interaction between quantum clocks.

\section{Two clocks}
Consider two gravitationally interacting clocks, labeled by $A$ and $B$, separated by a coordinate distance $x$ (in the frame of the far-away observer). To a lowest approximation to the solution of the Einstein equations, the gravitational interaction is described by the Newtonian gravitational energy $U(x) = - G m_A m_B/x$. In this work we focus only on the first order approximation to the solution for the metric. However, Post-Newtonian corrections can be analysed in the same manner. The labels $m_A$ and $m_B$ refer to the masses of particles $A$ and $B$, respectively. By these masses we mean the whole mass-energy contribution to the gravitational field, including both static rest mass $m$ and the dynamical mass, corresponding to the energy of the internal degrees of freedom $\hat{H}_{int}$. This notion of dynamical mass is of a purely relativistic nature, and arises from the interaction of the constituents of our composite particle. In fact, from a relativistic point of view, there is conceptually no difference between mass and interaction energy, and their distinction is effectively a matter of the energy scale with which the system is probed. The interaction can be described in quantum mechanical terms by promoting the masses of each particle to operators and using the mass-energy equivalence: $m\longrightarrow m + \hat{H}_{int}/c^2 $. For reasons of simplicity, we assume that the static mass is negligibly small as compared to the dynamical one and focus only on the effect due to the internal degrees of freedom. Thus, the Hamiltonian for the two-clock system is
\begin{equation}
\label{Hamiltonian2}
\hat{H} = \hat{H}_A + \hat{H}_B -\frac{G}{c^4x} \hat{H}_A \hat{H}_B.   
\end{equation} 
A full derivation regarding how the internal degree of freedom of a quantum particle evolves in perturbative general relativity is given in \cite{magdalena, igor}. For a heuristic discussion of discussion of Eq. (\ref{Hamiltonian2}) based on the superposition principle and the mass-energy equivalence, see Appendix (B). The same Hamiltonian can be obtained from a field theory perspective by the restriction to the two-particle sector of the field \cite{anastopoulos} and the use of the mass-energy equivalence, as we sketch in Appendix (C). Although the methods presented here suffice to describe the entanglement of clocks arising from gravitational interaction, a full description of the physics with no background spacetime would require a fundamental quantum theory of gravity. In the works of Rovelli \cite{rovelli} and Isham \cite{isham}, for example, it is suggested that time itself emerges from the dynamics of more fundamental degrees of freedom.
 
Let us assume that the energies of both Hamiltonians $\hat{H}_A$ and $\hat{H}_B$ are equal and that the initial state of the clocks is uncorrelated:
$\ket{\psi_{in}} = \left[(\ket{0} + \ket{1})/\sqrt{2}\right]^{\otimes 2}$.
The state at time $t$ according to the far-away observer is 
\begin{equation}
\label{superposition of time dilation}
\ket{\psi} = \frac{1}{\sqrt{2}}\left( \ket{0}\ket{\varphi_0} + \e^{-\frac{\mathrm{i} t}{\hbar}\Delta E}\ket{1}\ket{\varphi_1}\right),   
\end{equation}
where 
$\ket{\varphi_0} = \left(\ket{0}+\e^{-\frac{\mathrm{i} t}{\hbar}\Delta E}\ket{1}\right)/\sqrt{2}$, 	
and 
$\ket{\varphi_1} = \left(\ket{0}+\e^{-\frac{\mathrm{i} t}{\hbar}\Delta E\left(1-\frac{G \Delta E}{c^4 x}\right)}\ket{1}\right)/\sqrt{2}$.

We see from Eq. (\ref{superposition of time dilation}) that the clocks get entangled through gravitational interaction: the rate at which time runs in one clock is correlated to the value of the energy of the other clock. The state gets maximally entangled for the time
$
t_{mix} = \pi \hbar c^4 x/(G (\Delta E)^2).
$   
Using dimensionless variables in Planck units, $\tau = t/t_P$, $\varepsilon = \Delta E/E_P$, $\xi = x/l_P$, where $l_P = \sqrt{\hbar G/c^3}$ is the Planck length, $t_P = l_P/c$ is the Planck time, and $E_P = \hbar/t_P$ is the Planck energy, this time is expressed as
$
\tau_{mix} = \pi\xi/\varepsilon^2,    
$
in Planck time units.
As we approach $t_{mix}$, the reduced state of any of the clocks approaches the maximally mixed state and the clock is no longer able to function as a proper clock, since when we `ask the clock for the time' we get only random answers. Note that the presence of a static mass in the Hamiltonian would not alter the value of $t_{mix}$, as it would not enter in any part of the Hamiltonian that contributes to the entanglement between clocks. Specifically, the terms $m_A \mathbb{1}\otimes \hat{H}_B$, $\hat{H}_A \otimes m_B \mathbb{1}$ and $m_A \mathbb{1}\otimes m_B \mathbb{1}$ do not create any entanglement between $A$ and $B$, and do not change $t_{mix}$, whereas the term $\hat{H}_A \otimes \hat{H}_B$ does.

It is important to point out that for this effect to arise it is crucial that we consider the internal energy of the clocks as a quantum operator, instead of just taking into account the expectation value of the energy, as is done in semi-classical gravity. To explain this point, let us describe the evolution of clock $B$ under the influence of clock $A$, but with $H_A$ replaced by its expectation value. We assume that the initial state of both clocks is $\ket{\psi_{in}} = \ket{\psi_{in}}_A\otimes\ket{\psi_{in}}_B$. Following \cite{magdalena,igor}, the evolution equation for clock $B$ is $\mathrm{i}\hbar \partial_t\ket{\psi}_B = \dot{\tau}H_B\ket{\psi}_B$, where $\dot{\tau}$ is the derivative of the proper time $\tau$ with respect to $t$. By taking the expectation value of $H_A$, we have $\dot{\tau} = 1+\langle H_A\rangle/(c^4x)$, to first order in $c^{-2}$. Therefore, the state of $B$ at time $t$ is $\ket{\psi}_B = \mathrm{exp}\left(-\mathrm{i}t\left( 1+\langle H_A\rangle/(c^4x)\right)/\hbar\right)\ket{\psi_{in}}_B$. Because the situation is symmetric between $A$ and $B$ we can apply the same argument for $A$ and obtain, after evolution, a joint state of the form $\ket{\psi} = \ket{\psi}_A\otimes\ket{\psi}_B$, where $\ket{\psi}_A$ has the same form of $\ket{\psi}_B$ but with the labels $A$ and $B$ interchanged.
Then we have shown that, in the semi-classical approach, the clocks do not get entangled and the only result is an overall time dilation of one clock due to the \textit{mean} energy of the other clock. We will see below that this situation is effectively recovered in the classical limit of clocks.

Note that after $t_{mix}$ the purity of the reduced system will increase again. This fact is a consequence of the unitarity of the evolution of the composite system. 

The effect presented here has a fundamental influence on the measurement of time that follows only from quantum mechanics and general relativity in the weak-field limit. It is independent of the usual argument concerning limitations of the measurability of spacetime intervals due to black hole formation \cite{amelino-camelia, gambini}. As we will see later, the effect is significant in a parameter regime which is occurs long before formation of black holes becomes relevant. In order to strengthen the effect we next consider $N+1$ gravitationally interacting clocks, for $N\gg 1$.      

\section{N+1 clocks}
\label{Nclocks}
Now suppose there are $N+1$ clocks contained in a region of space characterised by the coordinate distance $x$. This array of clocks constitute a reference frame in the sense discussed in the Introduction. We ask the question of how the functioning of a single clock is affected by the presence of the other $N$ clocks. To give a lower bound on the effect we can consider $x$ to be the largest coordinate distance between any pair of clocks and write a generalisation of the interacting Hamiltonian of the previous section:  
\begin{align}
	\label{NHamiltonian}
\hat{H} =& 
       \sum_{a=0}^N \hat{H}_a -\frac{G}{c^4x}\sum_{a<b}\hat{H}_a \hat{H}_b,
\end{align}    
where the indices $a$ and $b$ label each of the individual clocks. In this part we concentrate only on the interacting part of the Hamiltonian, since we wish to analyse the loss of coherence of the reduced state of a single clock. We therefore analyse the evolution in the interaction picture. For an initial state of the form 
$\ket{\psi_{in}} = \left[(\ket{0} + \ket{1})/\sqrt{2}\right]^{\otimes N+1}$,
the reduced state of the zeroth clock is  
\begin{equation}
\rho_0 = \frac{1}{2}
\begin{pmatrix}
1 && \left[\frac{1}{2}\left(1+\e^{-\mathrm{i}\frac{\tau \varepsilon^2}{\xi}}\right)\right]^N\\
\left[\frac{1}{2}\left(1+\e^{\mathrm{i}\frac{\tau \varepsilon^2}{\xi}}\right)\right]^N && 1
\end{pmatrix}.   
\end{equation}    
Interestingly, the time for maximal mixing is independent of $N$ and is equal to $\tau_{mix}$ from the two-clock case. However, coherence can be significantly reduced for times earlier than $\tau_{mix}$. To quantify this we use the visibility $V$, defined by (twice) the absolute value of the non-diagonal element of the density operator.

In our case
\begin{equation}
V  =  2\vert \left(\rho_0\right)_{12} \vert \nonumber 
        = \left[\frac{1}{2}\left(1+\cos\frac{\tau \varepsilon^2}{\xi}\right)\right]^N \nonumber 
        \approx
        \, 1-\left(\frac{\sqrt{N}\tau\varepsilon^2}{2 \xi}\right)^2 \nonumber 
        \approx \, \e^{-\left(\frac{\sqrt{N}\tau\varepsilon^2}{2 \xi}\right)^2},
\end{equation}  
for $\tau \ll 2\xi/(\sqrt{N}\varepsilon^2)$. From here we identify a decoherence time which is, back in the initial units,
\begin{equation}
\label{decoherence time}
t_d = \frac{2\hbar c^4 x}{\sqrt{N}G (\Delta E)^2}.    
\end{equation}  
This characterises the fundamental limit on the time after which quantum clocks lose their ability to measure time when their gravitational effects are taken into account.

\begin{figure}[h]

\centering
\includegraphics[scale=0.6]{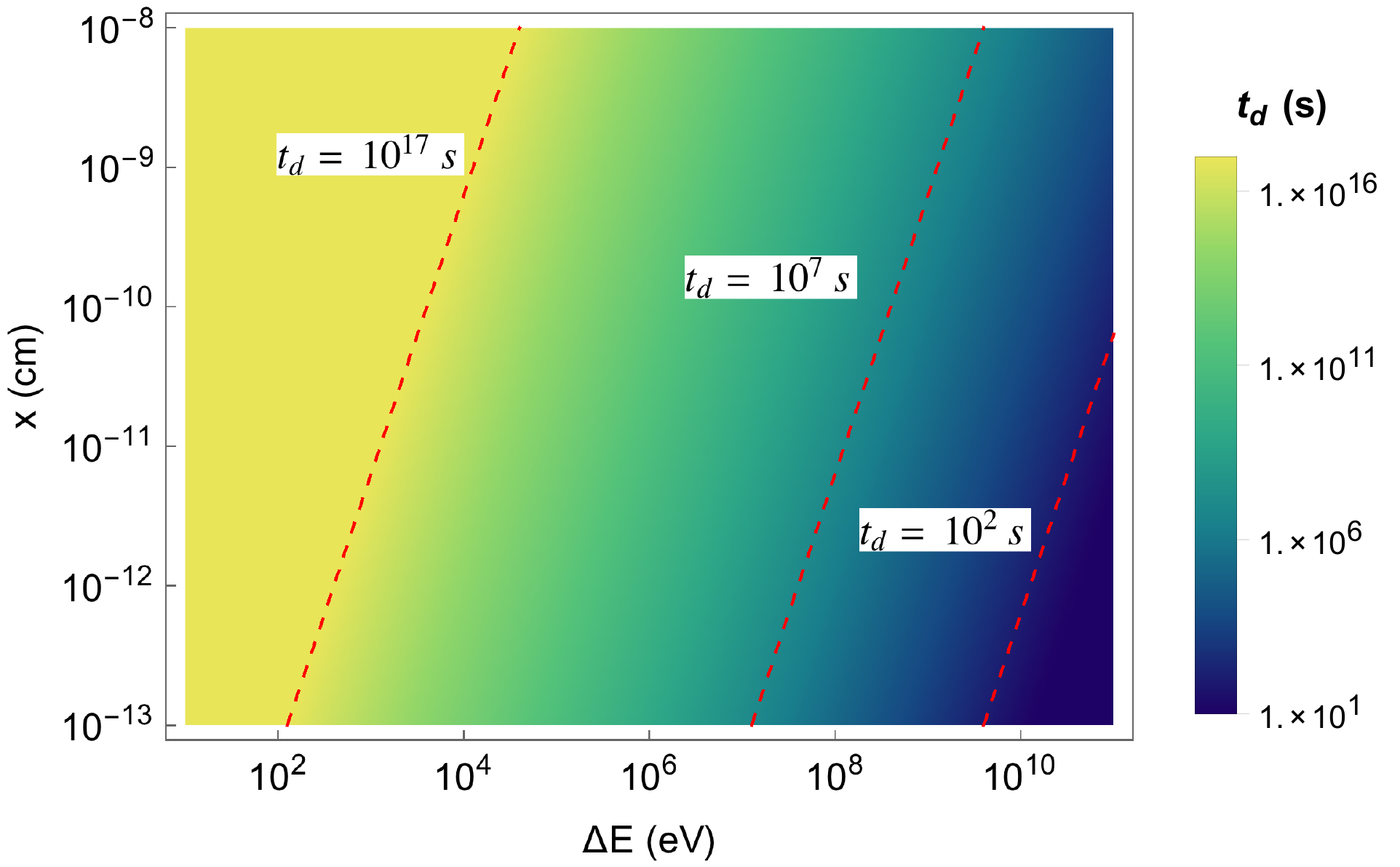}
\caption{\label{decoherencedensityplot} Clock decoherence time $t_d$ of Equation (\ref{decoherence time}) as a function of the clocks' energy gap ($\Delta E$) and the separation between clocks ($x$) for a macroscopic number of particles $N=10^{23}$. The dotted lines show three different decoherence time regimes for different scales of $\Delta E$ and $x$: $10^{17}$ s (the order of the age of the Universe), $10^{7}$ s (the order of one year) and $100$ s. Note that the blue region, showing relatively short decoherence times corresponds to energies and distances far from the Planck scale regime, suggesting a breakdown of the measurability of time at larger distance and lower energy scales.}
\end{figure}

We now give an estimate of the parameter regime where decoherence is significant. The calculations are done ignoring all effects external to our model and should be understood in terms of a \textit{gedanken}-experiment. The intention is to contrast the predictions given by our model with the usual predictions given by quantum gravity models, which do not expect limitations due to the combined effects of quantum mechanics and general relativity before the Planck scale. (For a discussion of the role of the Planck scale in the possibility of defining time, the reader may see \cite{amelino-camelia, padmanabhan, ng}). Figure (\ref{decoherencedensityplot}) shows the decoherence time $t_d$ as a function the energy gap $\Delta E$ and the distance $x$ for a macroscopic number of particles $N = 10^{23}$. Despite the fact that the effect is very small with respect to the regimes of current atomic clocks, it is important to analyse the order of magnitude of the limitations from a conceptual point of view. For instance, for a distance $x \approx 10^{-13}$ cm (the order of magnitude of the charge radius of a proton), an energy gap of $\Delta E \approx 10 \text{GeV}$, which is comparable to, for example, the energy of the nuclear bound state of a $K^-$ particle in $^4$He \cite{akaishi}, and a macroscopic number of particles $N \approx 10^{23}$, we find $t_d \approx 80 \text{s}$. The important point is that the regime of these parameters is several orders of magnitude away from the Plank scale. It is also important to note that, for these values of $\Delta E$, $N$, the Schwarzschild radius $r = 2GM/c^2$ (where $M = E/c^2$ and $E= N \Delta E$ is the total energy) is of the order of $10^{-29}$ m, so that the effect we predict is orders of magnitude away from the regime where a black hole is formed. 

To end this section we note that, despite the fact that this effect is not large enough to be measured with the current experimental capabilities, it might be possible to perform experiments on analogue systems to test this effect. Specifically, in Ref. \cite{katz} the authors consider an atom traversing an oscillating quantum reference frame, and show that the phase of the wave function of the atom has an uncertainty that can be related to the uncertainty in the atom's elapsed proper time. By the equivalence principle it is possible to interpret the acceleration that the oscillating reference frame induces on the atom as the gravitational effect that one clock suffers as a consequence of the presence of another nearby clock.

\section{Clocks in the classical limit}
\label{clocks as coherent states}
Given the ill-definedness of time measured by a single clock when it is in the presence of other clocks, how does the classical notion of a clock, including relativistic time dilation effects, arise? In what follows we answer this question by considering the classical limit of our model. The quantum state which is closest to the classical state of a clock is a spin or atomic coherent state. In general, spin coherent states can be defined as 
$
\vert \vartheta, \varphi, j \rangle = \left(\cos\frac{\vartheta}{2}\vert 0 \rangle +e^{i\varphi}\sin\frac{\vartheta}{2}\vert 1 \rangle \right)^{\otimes 2j} 	
$  
and can be understood in terms of a $j$-spin pointing in the direction given by the polar angles $\vartheta$ and $\varphi$. This picture is convenient since it  admits a Bloch sphere representation. We assume that the initial state of the clock is in a spin coherent state: 
$\ket{\psi_{in}} = \left[\left(\ket{0}+\ket{1}\right)\right/\sqrt{2}]^{\otimes 2j} 
                =\ket{\vartheta = \frac{\pi}{2}, \varphi = 0,j}.$

The Hamiltonian that evolves the state of this clock is the extension to angular momentum $j$ of the two-level ($j= 1/2$) Hamiltonian presented above. In terms of the angular momentum operator in the $z$ direction, $\hat{Z}$, Eq. (\ref{internal hamiltonian}) is written as $\hat{H}_{int} = \Delta E(\frac{1}{2}\mathbb{1}-\hat{Z})$. Therefore, for a spin-$j$ system the corresponding Hamiltonian is 
\begin{equation}
\label{coherent state hamiltonian}
\hat{H}_{free} = 
\Delta E(j \mathbb{1}-\hat{Z}), 
\end{equation}
where $\hat{Z} = \sum_{-j}^j  m \, \ketbra{m}{m}$. Note that the spectrum of this Hamiltonian is non-negative, ensuring the non-negativity of the mass when considering the mass-energy equivalence.
 
One of the approaches to the classical limit from within quantum mechanics is based on an experimental resolution that is coarse enough not to reveal the quantum features of the system \cite{kofler}. In our case we consider coarse-grained time measurements characterised by the experimental resolution $R$. The POVM corresponding to these measurements is defined by
$\left\{M_k\right\}_{k=1}^{2\pi/R}$, with
\begin{equation}
\label{POVM}
M_k = \frac{2j+1}{4\pi}\int_0^\pi\mathrm{d}\vartheta \sin\vartheta \int_{(k-1)R}^{k R}\mathrm{d}\varphi  \ketbra{\vartheta, \varphi}{\vartheta, \varphi}.      
\end{equation}
We can picture this POVM as dividing the Bloch sphere into $2\pi/R$ `bins'. The coherent state evolves by moving along the $\varphi$ direction, in the equator of the sphere. Then, the  probability that a measurement yields $k$ units, i.e. the probability for a spin coherent state $\rho$ to be in the $k$-th bin is
$p_k = \mathrm{Tr} M_k \rho 
    = \frac{2j+1}{4\pi}\int_0^\pi\int_{(k-1)R}^{k R}\mathrm{d}\vartheta\mathrm{d}\varphi \sin\vartheta \ Q_{\rho}(\vartheta,\varphi)$,
where $Q_{\rho}(\vartheta,\varphi) = \bra{\vartheta,\varphi}\rho\ket{\vartheta,\varphi}$ is the Husimi function, or $Q$-function, of the density matrix $\rho$. The characteristic width of this function is proportional to $j^{-1/2}$, and therefore, in the regime where $R\gg j^{-1/2}$ but still $j \gg 1$, the probability for the pointer of the clock to be found in more than one bin becomes negligible. Therefore, all of the fluctuations due to the quantum nature of the system are not visible in this regime and the clock behaves effectively classically. Note also that, after a coarse-grained measurement which finds the state of the clock in a particular bin, such state is effectively non-perturbed, since the part of it which lies outside the bin of size $R$ is negligible and therefore a projection on the region corresponding to the bin will not alter the state significantly. Therefore the clock will effectively continue its classical behaviour after measurement \cite{kofler2}.  
   
Consider now two clocks, labeled by $A$ and $B$, each of them being initially in a coherent state and interacting gravitationally with each other. For full generality we suppose that the $A$ ($B$) clock is a system with total spin $j_A$ ($j_B$). The full Hamiltonian is
\begin{equation}
\label{coherent_hamiltonian}
\hat{H} = \hat{H}_A + \hat{H}_B -\frac{G}{c^4 x}\hat{H}_A \hat{H}_B,    
\end{equation}
where $\hat{H}_A$ ($\hat{H}_B$) have the form (\ref{coherent state hamiltonian}).
For the initial state 
$\ket{\psi_{in}} = \ket{\vartheta = \frac{\pi}{2}, \varphi = 0,j_A}\otimes\ket{\vartheta = \frac{\pi}{2}, \varphi = 0,j_B}$,     
 the reduced state for the $B$ clock at time $t$ is
\begin{equation}
\label{reduced b state}
\rho_B = \frac{1}{4^{j_A}}\sum_{k=0}^{2j_A} \binom{2j_A}{k}\Big\vert \vartheta = \frac{\pi}{2},\varphi_k,j_B \Big\rangle \Big\langle \vartheta = \frac{\pi}{2},\varphi_k, j_B \Big\vert, 	
\end{equation}
where $\varphi_k = -\frac{t\Delta E}{\hbar}\left(1-\frac{G k \Delta E}{c^4 x}\right)$. Let us analyse closely this equation. It consists of a sum of coherent states, each of them evolving with a phase $\varphi_k$, modulated by a binomial distribution. The typical width of these coherent states depends on $j_B$. The state is measured by the coarse-grained POVM of Eq. (\ref{POVM}). When $j_B$ is large and $R \gg j^{-1/2}_B$, it is expected that each coherent state is significantly different from zero only inside one bin. This applies also for noninteracting clocks $A$ and $B$. However, in the presence of interaction the coherent states in the mixture of Eq. (\ref{reduced b state}) evolve with different time dilation factors, given by each of the phases $\varphi_k$. 

\begin{figure}
\centering
\includegraphics[scale=0.35]{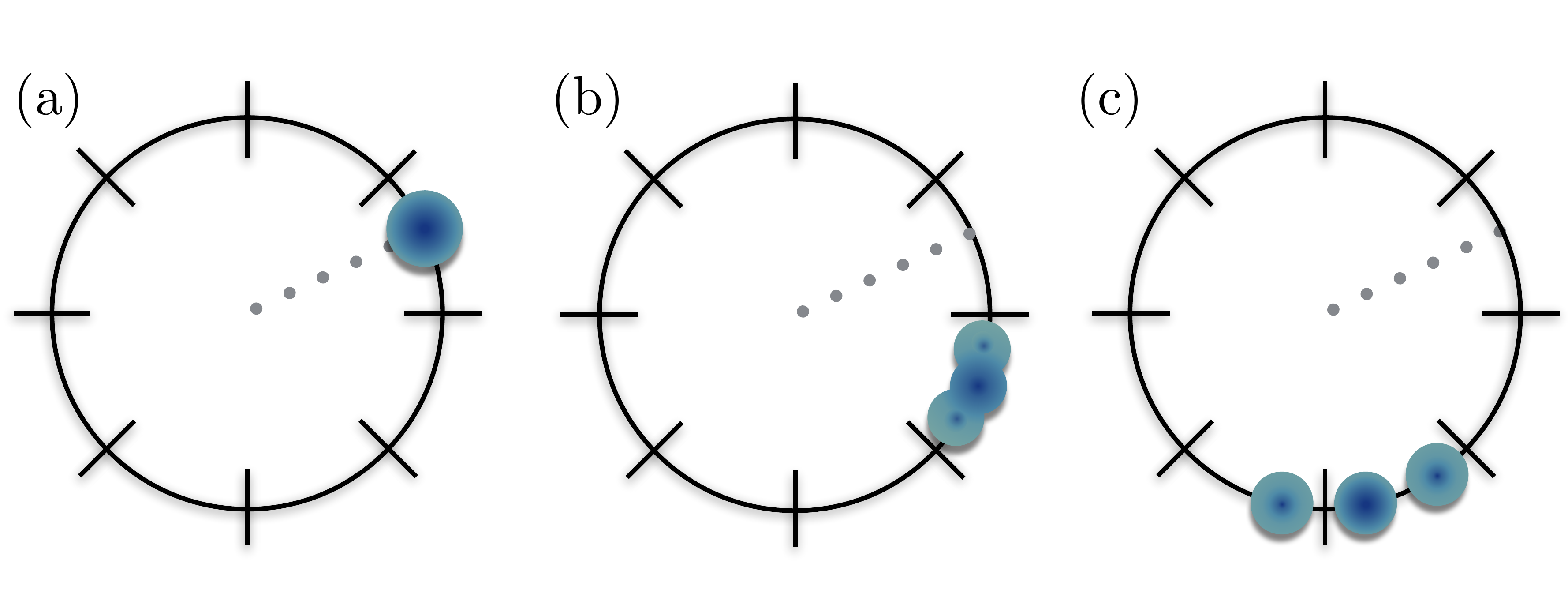}
\caption{\label{classical_limit} Emergence of the classical notion of general relativistic time dilation from the assumption of clocks in coherent states and coarse-grained measurements. Two coherent state clocks with spins $j_A$ and $j_B$ interact gravitationally. The reduced state for the $B$ clock is a sum of coherent states modulated by a binomial distribution (see Eq. (\ref{reduced b state})). The Husimi function of each coherent state is represented by a blue circle that precesses along the black circumference as it evolves. Each state in the mixture precesses at a different time dilation rate. The darker the colour of the circle representing the Husimi function at one location, the higher the probability of finding the clock pointer at that location. We consider a time measurement POVM that gives the probability of finding the clock state in one of the `bins' that divide the circle. The finite size of the bins represent the finite experimental resolution $R$. There are two effects that contribute to the classical limit: the average motion of the clock and the separation of the different Husimi functions. The average motion of the clock grows as $j_A$, while the effective separation of the coherent states is proportional to $\sqrt{j_A}$ (see Section (\ref{clocks as coherent states}) in the  main text). In picture \textbf{(a)} we depict the initial condition: all the Husimi functions are centred at the same point. As they evolve, the Husimi functions spread. For $t\ll t^* = \hbar c^4 x/(G \sqrt{2j_A} (\Delta E)^2)$ (picture \textbf{(b)}) and $R\gg j^{-1/2}_B$, all the coherent states lie inside one bin and no `quantum fluctuations' occur when measuring time. The $B$ clock is time-dilated according to the average energy of the clock $A$. For $t>t^*$, (picture \textbf{(c)}), different Husimi functions occupy different bins and the effects of the quantum entanglement between the clocks emerge despite the coarseness of the measurement.}
\end{figure}

There are two effects, different in nature, whose relative contributions to the evolution of $\rho_B$ give us the regime of parameters that defines the classical limit. The situation is depicted in Figure (\ref{classical_limit}). First, there is the `evolution of the clock as a whole', that is, the movement of the average phase of the clock $\varphi_{j_A} = -\frac{t\Delta E}{\hbar}\left(1-\frac{G j_A \Delta E}{c^4 x}\right)$. This phase, which corresponds to the pointer with the highest probability for detection, evolves time-dilated due to the average energy of the clock $A$. Second, coherent states tend to spread from each other, leading eventually to a mixing of the reduced state and therefore to ill-definedness of time measurements. To quantify these two effects we note, on the one hand, that the evolution of the time dilation part of the average phase is proportional to $j_A$. On the other hand, despite the fact that the angle separation of the coherent states $\Delta\varphi = \varphi_{2j_A}-\varphi_0 = \frac{2 G j_A(\Delta E)^2t}{c^4x\hbar}$ is also proportional to $j_A$, not all the terms in (\ref{reduced b state}) contribute significantly to the state, due to the binomial distribution $p(k) = 4^{-j_A}\binom{2j_A}{k}$. Indeed, for large $j_A$, $p(k)$ can be approximated by a Gaussian distribution, that is, 
$p(k) \approx \sqrt{\frac{1}{\pi j_A}}\exp\left(\frac{k-j_A}{\sqrt{j_A}}\right)^2$,
which has a characteristic width proportional to $\sqrt{j_A}$. This means that the \textit{effective} angle separation between coherent states grows with $\sqrt{j_A}$, rather than with $j_A$, that is, $\Delta \varphi_{eff} = \frac{G \sqrt{2j_A}(\Delta E)^2t}{\hbar c^4x}$. Therefore, for times much smaller than a characteristic time 
\begin{equation}
t^* = \frac{\hbar c^4 x}{G \sqrt{2j_A} (\Delta E)^2},
\end{equation}
say $t = \Gamma t^*$, where $\Gamma\ll 1$, the angle separation will grow as $\Delta \varphi_{eff} = \frac{t}{t^*} = \Gamma$, but we will have $\varphi_{j_A} = -\frac{c^4x }{G\Delta E\sqrt{2j_A}}\Gamma +\sqrt{\frac{j_A}{2}}\Gamma$, which grows as $\sqrt{j_A}$ for large values of $j_A$. Therefore, for this scaling with respect to $j_A$ and in the limit where $j_A\gg1$ we reach the regime where the classical limit of clocks holds, since entanglement is negligible at these scales. If apart from this characteristic time $t^*$ we have a coarse enough measurement, i.e. $R\gg j^{-1/2}_B$, measurements of time will detect time dilation, as classical general relativity predicts, but with no `quantum fluctuations'. Significantly, the time-dilation factor corresponds to the average energy of clock $A$, consistent with the semiclassical approximation to gravity in the quantum domain.

The evolution of the reduced state $\rho_B$ can also be studied in terms of its master equation. Following Ref. \cite{igor}, where a full treatment of the master equation for systems of particles evolving in the presence of relativistic time dilation is given, the master equation in this case can be written as   
\begin{equation}\label{masterequation}
\frac{\mathrm{d}\rho_B}{\mathrm{d}t} = \frac{\mathrm{i}}{\hbar} \left[ \hat{H}_B\left(1+\frac{G j_A \Delta E}{ c^4 x}\right), \rho_B \right] -\left(\sqrt{\frac{j_A}{2}}\frac{G\Delta E}{2c^4x}\right)^2\int_0^t\, \mathrm{d}s \Big[\hat{H}_B, \left[\hat{H}_B,\rho_B\right]_s\Big],	
\end{equation}
where  
$\left[\hat{H}_B,\rho_B\right]_s = e^{\frac{-\mathrm{i}s}{\hbar}\hat{H}_B}\left[\hat{H}_B,\rho_B\right]e^{\frac{\mathrm{i}s}{\hbar}\hat{H}_B}.$
We note that the first term, corresponding to the unitary part of the evolution, has a time dilation factor $\left(1+\frac{G j_A \Delta E}{ c^4 x}\right)$ that corresponds to the mean energy $j_A \Delta E$ of the $A$ clock. On the other hand, the second term, responsible for decoherence and quadratic in $\hat{H}_B$, is proportional to the square of the variance, $ j_A ( \Delta E )^2/2 $. We then see that, in a state of clock $A$ where the variance of the energy is negligible, clock $B$ evolves unitarily with a time-dilation factor given by the average energy of clock $A$, just as expected for a quantum state of matter in the semiclassical limit, where its energy-momentum tensor operator is replaced by its average value. For completeness we derive Eq. (\ref{masterequation}) in Appendix (D), following closely Ref. \cite{igor}. As seen in Appendix (D), the derivation of the master equation (\ref{masterequation}) holds in general for any quantum system and any form of the Hamiltonians $\hat{H}_A$ and $\hat{H}_B$. This fact implies that, as long as the initial state of the clocks is not in an energy eigenstate (a condition needed for the system to be a clock), the second term in Eq. (\ref{masterequation}) will be non-zero, as the variance of the energy will not vanish; implying that, irrespectively of the nature of the clocks, they will get entangled. 

Finally, in the light of the analysis of the present Section, let us now return to Eq. (\ref{clocks uncertainty relation}), obtained via a heuristic semi-classical argument in Section (\ref{clockmodel}), and show that it can also be derived from the classical limit of two interacting clocks, connecting the heuristic arguments based on the superposition principle and gravitational time dilation to our treatment of interacting clocks in the classical limit. Consider the two-clock scenario of the beginning of this Section with $j_A = 1/2$ and $j_B \gg 1$. We will analyse the limit in which the time-dilation of clock $B$ due to clock $A$ is significant (that is $G \Delta E_A/(c^4 x)$ is non-negligible), but the time-dilation effect on $A$ due to $B$ can be neglected, i.e. $G j_B \Delta E_B/(c^4 x) \ll 1$. Let us focus first on clock $B$. Its reduced state after evolution is given by $\rho_B = \frac{1}{2}(\ketbra{\vartheta = \frac{\pi}{2},\varphi_0, j_B}{\vartheta = \frac{\pi}{2},\varphi_0, j_B} + \ketbra{\vartheta = \frac{\pi}{2},\varphi_1, j_B}{\vartheta = \frac{\pi}{2},\varphi_1, j_B})$, with $\varphi_k = -\frac{t\Delta E_B}{\hbar}\left(1-\frac{G k \Delta E_A}{c^4 x}\right)$, $k = 0, \ 1$. We now define the operator 
\begin{equation}
\label{clock pointer}
T^{j_B} = \frac{\hbar(2j_B+1)}{4\pi\Delta E_B}\int_0^{\pi} \mathrm{d}\theta \sin\theta \int_0^{2\pi} \mathrm{d}\phi \, \phi \, \ketbra{\theta, \phi, j_B}{\theta, \phi, j_B}. 
\end{equation}
Physically, this operator represents the pointer position of clock $B$ and has physical dimensions of time. In the limit $j_B\longrightarrow \infty$, spin coherent states are orthonormal and therefore, the state $\ket{\vartheta = \frac{\pi}{2},\varphi_k, j_B}$ becomes an eigenstate of $T^{j_B}$ with eigenvalue $\hbar\varphi_k/\Delta E_B$, for $\varphi_k \in (0,2\pi)$ and $k = 0, \ 1$. Using this fact it is easy to show that, in this limit, the variance of the operator $T^{j_B}$ is given by
\begin{equation}
\Delta T^{j_B} = \frac{\hbar}{2\Delta E_B}(\varphi_1-\varphi_0) = \frac{G \Delta E_A t}{2 c^4 x}.	
\end{equation}
On the other hand, the probability of measuring one unit of time on clock $A$ is given by the operator $T_{A} = \frac{\hbar}{\Delta E_A} \ketbra{-}{-}$. Operationally, the time it takes for the average of $T_A$ to change significantly is given by $\mathrm{d}T_A = \Delta T_A/(\vert \mathrm{d}\langle T_A \rangle/\mathrm{d}t\vert)$, where the bars denote absolute value. We can now compute $\mathrm{d}T_A$ for the reduced state of clock $A$,  $\rho_A = \frac{1}{4^{j_B}}\sum_{k=0}^{2j_B} \binom{2j_B}{k}\Big\vert \vartheta = \frac{\pi}{2},\varphi_k,1/2 \Big\rangle \Big\langle \vartheta = \frac{\pi}{2},\varphi_k, 1/2 \Big\vert$, with $\varphi_k = -\frac{t\Delta E_A}{\hbar}\left(1-\frac{G k \Delta E_B}{c^4 x}\right)$. Since by assumption the time dilation effect of clock $B$ on clock $A$ is negligible, we take into account only the $\varphi_0$ contribution to $\mathrm{d}T_A$, yielding he result $\mathrm{d}T_A = \hbar/\Delta E_A$. Putting the pieces together we get
\begin{equation}
\mathrm{d}T_A \Delta T^{j_B} = \frac{\hbar G  t}{2 c^4 x},
\end{equation} 
which coincides with Eq. (\ref{clocks uncertainty relation}) up to a factor of $\pi/2$.

\section{Discussion}
In the (classical) picture of a reference frame given by general relativity, an observer sets an array of clocks over a region of a spacial hypersurface. These clocks trace world lines and tick according to the value of the metric tensor along their trajectory. Here we have shown that, under an operational definition of time, this picture is untenable. The reason does not only lie in the limitation of the accuracy of time measurement by a single clock, coming from the usual quantum gravity argument in which a black hole is formed when the energy density employed to probe spacetime lies inside the Schwarzschild radius for that energy. Rather, the effect we predict here comes from the interaction between nearby clocks, given by the mass-energy equivalence, the validity of the Einstein equations and the linearity of quantum theory. We have shown that clocks interacting gravitationally get entangled due to gravitational time dilation: the rate at which a single clock ticks depends on the energy of the surrounding clocks. This produces a mixing of the reduced state of a single clock, with a characteristic decoherence time after which the system is no longer able to work as a clock. Although the regime of energies and distances in which this effect is considerable is still far away from the current experimental capabilities, the effect is significant at energy scales which exist naturally in sub-atomic particle bound states. 

These results suggest that, in the accuracy regime where the gravitational effects of the clocks are relevant, time intervals along nearby world lines cannot be measured with arbitrry precision, even in principle. This conclusion may lead us to question wether the notion of time intervals along nearby world lines is well-defined. Because the spacetime distance between events, and hence the question wether the events are space-like, light-like or time-like separated, depend on the measurability of time intervals, one can expect that the situations discussed here may lead to physical scenarios with indefinite causal structure \cite{oreshkov}. The notion of well-defined time measurability is obtained only in the limit of high dimensional quantum systems subjected to accuracy-limited measurements. Moreover, we have shown that our model reproduces the classical time dilation characteristic of general relativity in the appropriate limit of clocks as spin coherent states. This limit is consistent with the semiclassical limit of gravity in the quantum regime, in which the energy-momentum tensor is replaced by its expectation value, despite the fact that in general the effect cannot be understood within this approximation.

The operational approach presented here and the consequences obtained from it suggest considering clocks as real physical systems instead of idealised objects might lead to new insights concerning the phenomena to be expected at regimes where both quantum mechanical and general relativistic effects are relevant.

\section{Acknowledgements}
We thank F. Costa, A. Feix, P. Hoehn, W. Wieland, and M. Zych for interesting discussions. We acknowledge support from the John Templeton Foundation, Project 60609, “Quantum Causal Structures”, from the research platform “Testing Quantum and Gravity Interface with Single Photons” (TURIS), and the Austrian Science Fund (FWF) through the special research program “Foundations and Applications of Quantum Sci- ence” (FoQuS), the doctoral program “Complex Quantum Systems” (CoQuS) under Project W1210-N25, and Individual Project 24621.

\appendix

\section{Analysis of the coordinate $t$}
\label{supplement1}
In this section we give an operational definition of the time coordinate $t$ introduced in the main text. It is defined as the proper time measured by an observer who is sufficiently far away from the clocks, such that the different states of his/her local clock, corresponding to different energy configurations of the clocks under study, are almost overlapping. In other words, the coordinate distance from the observer to the clocks is such that the different metrics originating from different energy eigenstates of the clocks are operationally indistinguishable at his/her location. The analysis given here considers only two clocks, but its generalisation to an arbitrary number of clocks is straightforward.

To give an estimate of such distance, consider that the observer measures time with a spin-$j$ coherent state clock, labeled by $C$, situated at a (finite) coordinate distance $r$ from the clocks under study, labeled by $A$ and $B$. We assume that $r\gg x$, where $x$ is the coordinate distance between $A$ and $B$, and compare the state of the observer's clock in the two extreme situations, i.e. when both $A$ and $B$ are in the ground state and when both are in the excited state. In the first case, the state of $C$ at time $t$ is
$
\ket{\psi_{00}} = \ket{\vartheta = \frac{\pi}{2}, \varphi = -\frac{t \varepsilon}{\hbar}},
$
where $t$ is the time measured by a clock that is far away enough so that the gravitational field is effectively zero at its location, and $\varepsilon$ is the energy gap of the clock $C$. (For simplicity we put $E_0 = 0$, and therefore $E_1 = E_1-E_0 = \Delta E$, as in the main text.) In the second case, where both $A$ and $B$ are in an excited state, the state of $C$ is $\ket{\psi_{11}} = \ket{\vartheta = \frac{\pi}{2}, \varphi = -\frac{t \varepsilon}{\hbar}(1-\frac{2G\Delta E}{c^4 r})}$. The overlap between these two states is
\begin{align}
\vert \brk{\psi_{00}}{\psi_{11}}\vert^2 = \frac{1}{4^j}\left(1+\cos\frac{2 G \Delta E \varepsilon t }{\hbar c^4 r}\right)^{2j}  
\approx  1-2j\left( \frac{2 G \Delta E \varepsilon t }{\hbar c^4 r} \right)^2.	
\end{align}
Consider now a finite measurement accuracy (i.e. a finite capability of distinguishing two quantum states) given by $\delta$, defined in such a way that that if $\vert \brk{\psi_{00}}{\psi_{11}}\vert^2 > 1-\delta$ then both states are effectively undistinguishable. Given this accuracy, indistinguishability is achieved for a distance $r$ satisfying
\begin{equation}
r > \frac{2\sqrt{2 j} G \Delta E \varepsilon t }{\hbar c^4 \sqrt{\delta}}.	
\end{equation}
For these distances we can use the label $t$ for the time coordinate in both cases. This procedure gives an operational meaning to coordinate time.

\section{Heuristic derivation of the two-clock Hamiltonian from superposition principle and mass-energy equivalence}
\label{supplement2}
In the following we discuss how the Hamiltonian for the evolution of an internal degree of freedom in a fixed static background [15] can be generalised to the case where the background is not fixed but, rather, is set by a quantum superposition of energies. Our basic assumptions are that the quantum mechanical principle of superposition and the laws of general relativity are valid. If this were not the case, we would be forced to conclude that at least one of the theories breaks down in this regime and new physics should emerge. We proceed iteratively by considering first the two-clock case and then extending to the three-clock case. The generalisation to higher number of clocks is then straightforward. 

 Consider two quantum clocks, labeled by $A$ and $B$, to be in the general state 
\begin{equation}
\ket{\psi_{in}} = (\alpha \ket{0}_A+\beta\ket{1}_A)\ket{\psi}_B.
\end{equation}
From Ref. [15] we know that for the amplitude $\alpha$, $\ket{\psi}_B$ evolves in the background produced by the state $\ket{0}_A$. Similarly, for the amplitude $\beta$, $\ket{\psi}_B$ evolves in the background produced by the state $\ket{1}_A$. Focusing on $\ket{\psi}_B$, without loss of generality, the evolution is given by
\begin{equation}
\ket{\psi(t)}_B = e^{-\frac{\mathrm{i}t}{\hbar}\dot{\tau}\hat{H}_B}\ket{\psi}_B,
\end{equation}
where $\hat{H}_B$ is the Hamiltonian of the internal degree of freedom of particle $B$ and $\dot{\tau}$ is the derivative of the proper time $\tau$ with respect to the coordinate time $t$. The operational meaning of $t$ is discussed in Appendix (\ref{supplement1}) above.  In the lowest-order approximation to the solution for the metric, we can write [12,15]
\begin{equation}
\dot{\tau}\approx 1+\frac{\Phi(x)}{c^2},	
\end{equation}
where $\Phi(x) = 0$ for the state $\ket{0}_A$ and $\Phi(x) = -G\Delta E/(c^2x)$ for the state $\ket{1}$.

In this way, $\ket{\psi}_B$ evolves as 
\begin{align}
\ket{\psi}_B \longrightarrow & e^{-\frac{\mathrm{i}t}{\hbar}\hat{H}_B}\ket{\psi}_B \ \ \text{for} \ \ \ket{0}_A,\\
\ket{\psi}_B \longrightarrow & e^{-\frac{\mathrm{i}t}{\hbar}\hat{H}_B(1-\frac{G\Delta E}{c^4x})}\ket{\psi}_B \ \ \text{for} \ \ \ket{1}_A.	
\end{align}
The phases in the previous equations already include the gravitational interaction between particles $A$ and $B$, as well as the free evolution of $B$. Therefore, the evolution of $\ket{0}_A$ ($\ket{1}_A$) is merely given by the phase corresponding to $E_0$ ($E_1$), that is, $\ket{0}_A\longrightarrow \ket{0}_A$ and $\ket{1}_A \longrightarrow e^{-\frac{\mathrm{i}t}{\hbar}\Delta E}\ket{1}_A$. 

Now, applying the superposition principle, we write the solution for the evolved state as a linear combination of the solutions for each energy:
\begin{equation}
\ket{\psi_{in}} \longrightarrow \alpha \ket{0}_A  e^{-\frac{\mathrm{i}t}{\hbar}\hat{H}_B}\ket{\psi}_B   
+\beta e^{-\frac{\mathrm{i}t}{\hbar}\Delta E} \ket{1}_A e^{-\frac{\mathrm{i}t}{\hbar}\hat{H}_B(1-\frac{G\Delta E}{c^4x})}\ket{\psi}_B = \ket{\psi_{fin}}.
\end{equation}
This evolution can be expressed in terms of the Hamiltonian (4), that is,
\begin{equation}
\ket{\psi_{fin}} = e^{-\frac{\mathrm{i}t}{\hbar}(\hat{H}_A + \hat{H}_B -\frac{G}{c^4x} \hat{H}_A \hat{H}_B)}\ket{\psi_{in}}.
\end{equation}

We now extend the result of the two-clock case to the case where we have three clocks, labeled by $A$, $B$ and $C$. We assume that the coordinate distance between each pair of particles is $x$, according to the observer far away. The idea is to single out one of the particles, say $A$, as the one whose energy eigenstates set the metric background, and use the result of the two particle case for the remaining particles. We write the initial state as
\begin{equation}
\ket{\psi_{in}} = (\alpha \ket{0}_A+\beta\ket{1}_A)\ket{\psi}_{BC}.	
\end{equation}
Following the steps for the two particle case, we write
\begin{equation}
\ket{\psi_{in}} \longrightarrow \alpha \ket{0}_A  e^{-\frac{\mathrm{i}t}{\hbar}H_{BC}}\ket{\psi}_{BC}   
	+\beta e^{-\frac{\mathrm{i}t}{\hbar}\Delta E} \ket{1}_A e^{-\frac{\mathrm{i}t}{\hbar}H_{BC}(1-\frac{G\Delta E}{c^4x})}\ket{\psi}_{BC} = \ket{\psi_{fin}},	
\end{equation}
where $H_{BC} = \hat{H}_B + \hat{H}_C -\frac{G}{c^4 x} \hat{H}_B \hat{H}_C$ is the joint Hamiltonian for the particles $B$ and $C$, derived from the analysis of two particles. Explicitly, the phase corresponding to the state $\ket{1}_A$ is 
\begin{align}
\frac{t}{\hbar} H_{BC}(1-\frac{G\Delta E}{c^4x}) = \frac{t}{\hbar} (\hat{H}_B+\hat{H}_C-\frac{G}{c^4x}(\hat{H}_B\hat{H}_C+ \Delta E \hat{H}_B+\Delta E \hat{H}_C) +\frac{G^2}{c^8x^2} \Delta E \hat{H}_B \hat{H}_C).
\end{align} 
Note that the last term in the last equation is higher order in $c^{-2}$ and therefore is neglected at the first level of approximation. Therefore we have
\begin{align}
\ket{\psi_{fin}} =& 	\alpha \ket{0}_A  e^{-\frac{\mathrm{i}t}{\hbar}\left(\hat{H}_B+\hat{H}_C-\frac{G}{c^4x}\hat{H}_B \hat{H}_C\right)}\ket{\psi}_{BC}   
	+\beta e^{-\frac{\mathrm{i}t}{\hbar}\Delta E} \ket{1}_A e^{-\frac{\mathrm{i}t}{\hbar}\left(\hat{H}_B + \hat{H}_C -\frac{G\Delta E}{c^4x}(\Delta E\hat{H}_B +\Delta E\hat{H}_C+ \hat{H}_B \hat{H}_C )\right)}\ket{\psi}_{BC} \\
	=& e^{-\frac{\mathrm{i}t}{\hbar}H_{ABC}}\ket{\psi_{in}},
\end{align}
where
\begin{equation}
H_{ABC} = \hat{H}_A+ \hat{H}_B +\hat{H}_C -\frac{G}{c^4x}(\hat{H}_A\hat{H}_B+\hat{H}_A\hat{H}_C+\hat{H}_B\hat{H}_C).	
\end{equation}
It is clear now that the generalisation for $N$ particles is given by the Hamiltonian
\begin{equation}
H = \sum_a \hat{H}_A -\frac{G}{c^4}\sum_{a<b} \frac{\hat{H}_A \hat{H}_B}{\vert x
_a-x_b\vert},	
\end{equation}
which reduces to Eq. (6) under the approximation $\vert x_a-x_b\vert \approx x$ for all $a$, $b$.

\section{Two clock Hamiltonian from Quantum Field Theory approach}
\label{supplementQFT}
The Hamiltonian (4) can also be obtained from a quantum field theory in the weak-field limit by restricting to the two particle subspace, as done in [16], and using the mass-energy equivalence. In the following we sketch a derivation of this fact using natural units ($c = 1$ and $\hbar = 1$). For a detailed presentation, the reader may consult [16]. We implement the mass energy equivalence, in the sense of the main text, from the beginning of our calculation by considering our composite particle to emerge from the interaction of two scalar fields $\varphi_1$ and $\varphi_2$. The Lagrangian density of the field coupled to gravity reads  
\begin{equation}
\label{lagrangian}
\mathcal{L} = - \frac{1}{2}\sqrt{-g}\left(\sum_Ag^{\mu\nu}\left(\partial_{\mu}\varphi_A\right)\left(\partial_{\nu}\varphi_A\right) + \sum_{AB} M^2_{AB}\varphi_A\varphi_B\right), 	
\end{equation}
where $g$ denotes the determinant of the metric $g_{\mu\nu}$ and $M_{AB}$ is a symmetric matrix that couples the fields $\varphi_1$ and $\varphi_2$ (c.f. \cite{peskin}). This interaction gives rise to the mass of the composite particle. We denote the eigenvalues of $M_{AB}$ by $m$ and $m+\Delta E$. It is important to stress that, as noted in the main text, there is fundamentally no difference between mass and interaction energy in a relativistic theory. The distinction between static mass and dynamical mass, i.e. the mass that arises from the interaction of internal degrees of freedom, is only an effective one, depending on the energy scale with which the system is probed. In this sense, the matrix $M_{AB}$ can be interpreted as a sum of static mass ($m$) contribution and internal energy or dynamical mass (with eigenvalue $\Delta E$) contribution. It provides an effective description of a composite particle with different energy levels. A full treatment of the dynamics of composite particles in quantum field theory is a research area on its own (see for example \cite{stumpf}) and is beyond the scope of our paper.

We write the Lagrangian density in the form (\ref{lagrangian}) to make explicit the fact that we will take superpositions of different energy eigenstates of the internal Hamiltonian, in the sense of the main text. Since $M_{AB}$ is symmetric, there exists an orthogonal matrix $C_{AB}$ that diagonalises it. 
We assume a metric field in the weak field limit $\mathrm{d}s^2 = -(1+2\Phi(\mathbf{x}))\mathrm{d}t^2+\mathbf{\mathrm{d}x \cdot \mathrm{d}x}$, and calculate, via a Legendre transformation, the Hamiltonian in this approximation
\begin{equation}
\label{fieldhamiltonian}
H = \frac{1}{2} \int \mathrm{d}^3\mathbf{x} \left(1+\Phi(\mathbf{x})\right)\left(\sum_A \left(\pi_A^2 +(\nabla\varphi_A)^2 \right) + \sum_{AB}M^2_{AB}\varphi_A \varphi_B\right).
\end{equation}
Here $\pi_A = \dot{\varphi_A} $ denotes the canonical conjugate momentum to $\varphi_A$. The gravitational potential $\Phi$ satisfies the equation $\nabla^2\Phi = -4\pi \rho$, where $\rho = \sum_A \left(\pi_A^2 +(\nabla\varphi_A)^2 \right) + \sum_{AB}M^2_{AB}\varphi_A \varphi_B$ is the energy density of the matter field.

In order to quantise the field, we first write the Hamiltonian (\ref{fieldhamiltonian}) in the basis in which $M_{AB}$ is diagonal,
$H = \frac{1}{2} \sum_A \int \mathrm{d}^3\mathbf{x} \left(1+\Phi(\mathbf{x})\right)\left(p_A^2 +(\nabla\Psi_A)^2  + \mu^2_{A}\Psi_A^2\right)$,
where $\Psi_A = \sum_B C_{AB} \varphi_B$, $C_{AB}$ is the matrix that diagonalises $M_{AB}$, and $p_A$ is the momentum conjugate to $\Psi_A$. The matrix $M_{AB}$ has eigenvalues $\mu_{A}$, for $A = 1$, $2$.  

We then Fourier-expand $\Psi$ and $p_A$
\begin{subequations}
\label{fourierpsi}
\begin{align}
\Psi_A(\mathbf{x}) &= \int\frac{\mathrm{d}^3\mathbf{k}}{(2\pi)^3\sqrt{2\omega_{A,\mathbf{k}}}}\left(\mathrm{e}^{-\mathrm{i}k_Ax}b_{A,\mathbf{k}}+\mathrm{e}^{\mathrm{i}k_Ax}b^\dagger_{A,\mathbf{k}}\right) \\
p_A(\mathbf{x})&= \mathrm{i}\int\frac{\mathrm{d}^3\mathbf{k}}{(2\pi)^3}\sqrt{\frac{\omega_{A,\mathbf{k}}}{2}}\left(\mathrm{e}^{-\mathrm{i}k_Ax}b_{A,\mathbf{k}}+\mathrm{e}^{\mathrm{i}k_Ax}b^\dagger_{A,\mathbf{k}}\right), 	
\end{align}
\end{subequations}
where $k_Ax = -\omega_{A,\mathbf{k}} t + \mathbf{k}\cdot\mathbf{x}$, and $\omega_{A,\mathbf{k}}^2 = \mathbf{k}^2 + \mu_A^2$, and impose the commutation relations 
$\left[b_{A,\mathbf{k}},b^\dagger_{\mathbf{A^\prime k^\prime}}\right] = \delta_{A,A^\prime}\delta^3(\mathbf{k}-\mathbf{k^\prime})$
Next we insert the expressions (\ref{fourierpsi}) in the Hamiltonian and then take the slow velocity approximation of the fields
\begin{subequations}
\label{nrapprox}
\begin{align}
\Psi_A(\mathbf{x})\approx & \frac{1}{\sqrt{2\mu_a}}\left(\chi_A(\mathbf{x})+\chi_A^\dagger(\mathbf{x})\right) \\
p_A(\mathbf{x}) \approx & \, \mathrm{i}\,\sqrt{\frac{\mu_a}{2}}\left(\chi_A(\mathbf{x})-\chi_A^\dagger(\mathbf{x})\right),  	
\end{align}	
\end{subequations} 
where $\chi_A(\mathbf{x}) = (2\pi)^{-3}\int \mathrm{d}^3\mathbf{k}\mathrm{e}^{\mathrm{i}(\mu_At-\mathbf{k}\cdot \mathbf{x})}b_{A,\mathbf{k}}$. Then we solve the Poisson equation for the potential
\begin{equation}
\label{gravitationalpotential}
\Phi(\mathbf{x}) = -G \int \mathrm{d}^3\mathbf{x^\prime}\frac{\rho(\mathbf{x^\prime})}{\vert \mathbf{x}-\mathbf{x^\prime}\vert},
\end{equation}
insert the expression (\ref{gravitationalpotential}) into (\ref{fieldhamiltonian}), and keep terms within the slow velocity approximation. The (normal-ordered) Hamiltonian then reads 
\begin{equation}
\label{matrixhamiltonian}
H = \sum_{AB}\int \mathrm{d}^3\mathbf{x} \left(M_{AB} \phi^\dagger_A \phi_B- \frac{1}{2} M^{-1}_{AB} \phi^\dagger_A\nabla^2\phi_B\right) -G	\sum_{ABCD}\int  \mathrm{d}^3\mathbf{x}{d}^3\mathbf{x^\prime} \frac{(M_{AB}\phi^\dagger_A\phi_B)(M_{CD} \phi^\dagger_C \phi_D)}{\vert \mathbf{x}-\mathbf{x^\prime}\vert},
\end{equation}
where $\phi_A = \sum_B C^{-1}_{AB}\chi_{B}$.
The expression $ \rho(\mathbf{x}) = \sum_{AB} M_{AB} \phi^\dagger_A(\mathbf{x}) \phi_B(\mathbf{x})$ has the interpretation of a mass-energy density of the field. However, since it involves the product of two field operators at the same point, it is not a well-defined operator and leads to divergencies. This problem is handled by a suitable regularisation procedure and leads to a renormalisation of the mass [16]. The regularised mass density is given by $ \rho_{reg}(\mathbf{x}) = \sum_{AB} \int \mathrm{d}^3\mathbf{x^\prime} f_{\delta} (\mathbf{x}-\mathbf{x^\prime})M_{AB} \phi^\dagger_A(\mathbf{x^\prime}) \phi_B (\mathbf{x^\prime})$, where $ f_{\delta}$ is a normalised, positive function dependent on a regularisation parameter $\delta$ in such a way that $f_{\delta}(\mathbf{x})\longrightarrow\delta^3(\mathbf{x})$ as $\delta\longrightarrow0$.

In order to obtain the restriction of (\ref{matrixhamiltonian}) to the two-particle sector of the Fock space, we compute the matrix element $\bra{\xi^{(1)},\eta^{(1)}} H \ket{\xi^{(2)},\eta^{(2)}}$, where $\ket{\xi^{(i)},\eta^{(i)}} = 2^{-1/2}\sum_{AB}\int \mathrm{d}^3\mathbf{x}{d}^3\mathbf{x^\prime} \xi_A (\mathbf{x}) \eta_B(\mathbf{x^\prime})\phi^\dagger_A(\mathbf{x})\phi^\dagger_B(\mathbf{x^\prime})\ket{0}$ is a two-particle state for $i = 1,2$ (here $\ket{0}$ denotes the vacuum state of the field). From this matrix element we can then read off the form of the two particle Hamiltonian, which we can write as
\begin{equation}
\label{twoparticlehamiltonian}
\hat{H} = \hat{M}_{ren}\otimes\mathbb{1}+\mathbb{1}\otimes\hat{M}_{ren} + \frac{1}{2}\hat{M}^{-1}\hat{\mathbf{p}}^2\otimes\mathbb{1}+ \frac{1}{2}\hat{M}^{-1}\mathbb{1}\otimes\hat{\mathbf{p}}^2-G\frac{\hat{M}\otimes\hat{M}}{\vert \hat{\mathbf{x}}\otimes\mathbb{1}-\mathbb{1}\otimes\hat{\mathbf{x}} \vert},	
\end{equation}
where $\bra{\xi}\hat{M}\ket{\eta} = \sum_{AB}\int \mathrm{d}^3\mathbf{x} \bar{\xi}_A(\mathbf{x})M_{AB}\eta_B(\mathbf{x})$, and $\hat{M}_{ren} = \hat{M}- (\pi\delta^2)^{-1/2} G \hat{M}^2 $ is the renormalised mass matrix. The Hamiltonian (\ref{twoparticlehamiltonian}) is equivalent to the Hamiltonian (4) together with the kinetic part, which we have ignored in the main text. The Hamiltonian for an arbitrary number of particles can be obtained from this approach by projecting in the corresponding subspace. In the same way as for the matrix $M_{AB}$, the operator $\hat{M}$ can be interpreted as a sum of static mass and internal energy.

\section{Derivation of the Master Equation}
\label{supplement3}
In this section we derive the master equation for the evolution of a single coherent state clock, labeled by $B$, interacting gravitationally with another coherent state clock, labeled by $A$. We follow closely the derivation presented in Ref. [15]. The Hamiltonian of the system is given by Eq. (11), which we write here as
\begin{align}
\hat{H} =& \hat{H}_A+\hat{H}_B + \lambda\hat{H}_A\hat{H}_B \\
 =& \hat{H}_0 +\hat{H}_{int},
\end{align}
where $\hat{H}_0 = \hat{H}_A+\hat{H}_B$ and $\hat{H}_{int} = \lambda\hat{H}_A\hat{H}_B$. In our case $\lambda = -G/(c^4x)$, but the derivation is completely general. We assume that the state at $t = 0$ is uncorrelated: $\rho(0) = \rho_A(0)\otimes\rho_B(0)$. The evolution for the full state is given by
\begin{equation}
\mathrm{i}\hbar \dot{\rho}(t) = [\hat{H}, \rho(t)].	
\end{equation}
For a general operator $\hat{A}$, we define
\begin{equation}
\tilde{A} = e^{\frac{\mathrm{i}t}{\hbar}(\hat{H}_0 + \hat{h})} \, \hat{A} \, e^{-\frac{\mathrm{i}t}{\hbar}(\hat{H}_0 +\hat{h})},
\end{equation}
where $\hat{h} = \lambda \bar{E}_A\hat{H}_B$, and $\bar{E}_A = \mathrm{Tr} (\rho_A \hat{H}_A)$. Now we apply this transformation to the total density operator of the system. In terms of $\tilde{\rho}$, the equation of motion is
\begin{equation} \label{vonneumanntilde}
	\mathrm{i} \hbar \frac{\mathrm{d}\tilde{\rho}}{\mathrm{d}t}	= [\tilde{H}_{int}-\tilde{h},\tilde{\rho}].
\end{equation}
The implicit solution to this equation is 
\begin{equation} \label{implicitsolution}
	\tilde{\rho}(t) = \tilde{\rho}(0) -\frac{\mathrm{i}}{\hbar}\int_0^t \mathrm{d}s	[\tilde{H}_{int}-\tilde{h},\tilde{\rho}(s)].
\end{equation}
Substituting (\ref{implicitsolution}) into (\ref{vonneumanntilde}) yields
\begin{equation}
\frac{\mathrm{d}\tilde{\rho}}{\mathrm{d}t} = -\frac{\mathrm{i}}{\hbar}[\tilde{H}_{int}-\tilde{h},\tilde{\rho}(0)]-\frac{1}{\hbar^2}\int_0^t \mathrm{d}s	[\tilde{H}_{int}-\tilde{h},[\tilde{H}_{int}-\tilde{h},\tilde{\rho}(s)]].
\end{equation}
We now approximate the last equation to the second order in $\tilde{H}_{int}$ and replace $\tilde{\rho}(s)$ by $\rho_A(0)\otimes\tilde{\rho}_B(s)$. With this approximation we trace over the $A$ clock and obtain an equation for $\tilde{\rho}_B$:
\begin{align}
\frac{\mathrm{d}\tilde{\rho}_B}{\mathrm{d}t} \approx &  - \frac{\mathrm{i}}{\hbar}\mathrm{Tr}_A\left([\tilde{H}_{int}-\tilde{h},\rho_A(0)\otimes\tilde{\rho}_B(0)]\right)-\frac{1}{\hbar^2}\int_0^t \mathrm{d}s	\mathrm{Tr}_A\left([\tilde{H}_{int}-\tilde{h},[\tilde{H}_{int}-\tilde{h},\rho_A(0)\otimes\tilde{\rho}_B(s)]]\right)	 \\
 =& -\left(\frac{\lambda \Delta E_A}{\hbar}\right)^2\int_0^t\mathrm{d}s [\hat{H}_B,[\hat{H}_B,\tilde{\rho}(s)]],
\end{align} 
where we have defined $\Delta E_A = \sqrt{\mathrm{Tr}\left(\rho_A(\hat{H}_A-\bar{E}_A\mathbb{1}_A)^2\right)}$ and have taken into account the fact that $\hat{H}_B = \tilde{H}_B$. Changing back to the original density operator $\rho_B$ and writing the value of $\lambda$ explicitly, we find
\begin{equation}
\frac{\mathrm{d}\rho_B}{\mathrm{d}t} = \frac{\mathrm{i}}{\hbar} \left[ \hat{H}_B\left(1+\frac{G j_A \Delta E}{ c^4 x}\right), \rho_B \right] -\left(\sqrt{\frac{j_A}{2}}\frac{G\Delta E}{2c^4x}\right)^2\int_0^t\, \mathrm{d}s \Big[\hat{H}_B, \left[\hat{H}_B,\rho_B\right]_s\Big],	
\end{equation}
where
\begin{equation}
\left[\hat{H}_B,\rho_B\right]_s = e^{\frac{-\mathrm{i}s}{\hbar}\hat{H}_B}[\hat{H}_B,\rho_B]e^{\frac{\mathrm{i}s}{\hbar}\hat{H}_B},	
\end{equation}
which coincides with Eq. (14).

	




\end{document}